\documentclass{ocg} 
\usepackage{multirow}
\usepackage{listings} 
\usepackage{graphicx} 
\usepackage{caption}
\usepackage{enumerate}
\def\NN{\mathbb{N}}
\def\FF{\mathbb{F}}
\def\LL{\mathcal{L}}
\def\trans{M=\left<\mathcal{X},\mathcal{Y},S,\delta,\lambda\right>}
\DeclareMathOperator{\NM}{\mathsf{NM}}
\DeclareMathOperator{\TNM}{\mathsf{\overline{m}L}}
\DeclareMathOperator{\CT}{\mathsf{C}}
\DeclareMathOperator{\TT}{\mathsf{T}}
\DeclareMathOperator{\LT}{\mathsf{L}}
\DeclareMathOperator{\EC}{\mathsf{EC}}
\DeclareMathOperator{\TM}{\mathsf{mL}}
\DeclareMathOperator{\SNF}{SNF}

\DeclareMathOperator{\rank}{rank}
\newcommand{\FT}{FT}
\newcommand{\LFT}{LFT}

\lstdefinelanguage{pylang}{morekeywords={def,for,if,then,do,while,with,return,break,else,continue,and,nil,output,select,elif,in},morecomment=[l]{\%}}
 
 \DeclareCaptionFormat{listing}{\rule{\dimexpr\textwidth\relax}{0.4pt}\par\vskip1pt#1#2#3}
\begin{document}
\title{STATISTICAL STUDY ON THE NUMBER OF INJECTIVE LINEAR FINITE TRANSDUCERS}
\runningtitle{STATISTICAL STUDY ON THE NUMBER OF INJECTIVE LFTS}
\author[e]{Ivone Amorim} 
\author[z]{Ant\'onio
  Machiavelo} 
\author[e]{Rog\'erio Reis} 
\address{CMUP, Faculdade de
  Ci\^encias da Universidade do Porto, Portugal} 
\address[e]{
  \email{\{ivone.amorim,rvr\}@dcc.fc.up.pt}}
\address[z]{\email{ajmachia@fc.up.pt}}
\maketitle
\begin{abstract}
The notion of linear finite transducer (\LFT{}) plays a crucial role in some cryptographic systems. In this paper we present a way to get an approximate value, by random sampling, for the number of non-equivalent injective LFTs.   By introducing a recurrence relation to count canonical \LFT{}s, we show how to estimate the percentage of $\tau$-injective \LFT{}s. Several experimental results are presented, which by themselves constitute an important step towards the evaluation of the key space of those systems.
\end{abstract}
\thispagestyle{empty}
\section{Introduction}

In this work we present a statistical study on the number of
non-equivalent linear finite transducers that are injective with some delay. This study
is motivated by the application of these transducers in
Cryptography. A transducer, in this context, is a finite state
sequential machine given by a quintuple
$\left<\mathcal{X},\mathcal{Y},S,\delta,\lambda\right>$, where:
$\mathcal{X}$, $\mathcal{Y}$ are the nonempty input and output
alphabets, respectively; $S$ is the nonempty finite set of states;
$\delta: S\times \mathcal{X}\to S$, $\lambda: S\times \mathcal{X}\to
\mathcal{Y}$, are the state transition and output functions,
respectively. These transducers are deterministic and can be seen as
having all the states as final. Every state in $S$ can be used as
initial state, and this gives rise to a transducer in the usual sense,
\textit{i.e.}, one that realises a rational function. Therefore, in
what follows, a transducer is a family of classical transducers that
share the same underlying digraph.

A finite transducer is called linear if its transition and output functions
are linear maps.  Linear finite transducers play a core role in a family of
cryptosystems, named FAPKCs, introduced in a series of papers by
Tao~\cite{TaoChen1985,TaoChenChen1997,TaoChen1997,TaoChen1999}. Those
schemes seem to be a good alternative to the classical ones, being
computationally attractive and thus suitable for application on
devices with very limited computational resources, such as satellites,
cellular phones, sensor networks, and smart cards \cite{TaoChen1997}.

\begin{figure}[t]
  \begin{center}
  \includegraphics[width=14cm]{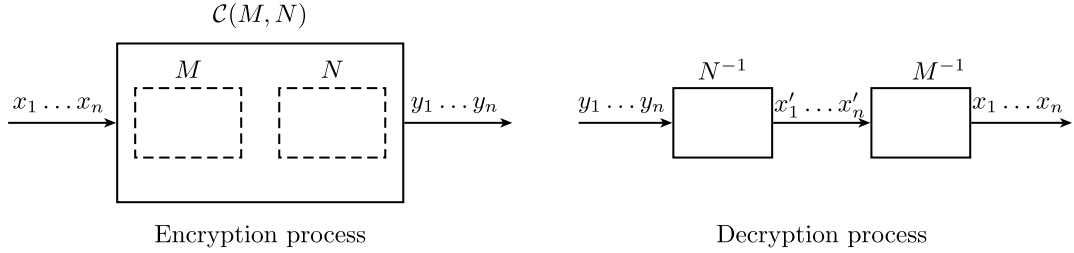}
  \end{center}
  \caption{Schematic representation of FAPKC working principle.}
  \label{fig1}
\end{figure}

Roughly speaking, in these systems, the private key consists of two
injective transducers, denoted by $M$ and $N$ in Figure~\ref{fig1},
where $M$ is a linear finite transducer (\LFT), and $N$ is a
non-linear finite transducer (non-\LFT{}) of a special kind, whose
left inverses can be easily computed. The public key is the result of
applying a special product, $\mathcal{C}$, for transducers to the
original pair, obtaining a non-\LFT{}, denoted by
$\mathcal{C}(M,N)$ in Figure~\ref{fig1}. The crucial point is that it
is easy to obtain an inverse of $\mathcal{C}(M,N)$ from the inverses
of its factors, $M^{-1}$ and $N^{-1}$, while it is believed to be hard
to find that inverse without knowing those factors.  On the other
hand, the factorization of a transducer seems to be a hard problem by
itself~\cite{Zongduo1998}.

\LFT{}s are fundamental in the FAPKC systems because their invertibility theory is of core importance in the security of these systems.
 They also play a crucial role in the key generation process, since in these systems a
pair $(\text{public key, private key})$ is formed using one injective \LFT{} and
two injective non-\LFT{}s, as explained above. Consequently, for these
cryptosystems to be feasible, injective \LFT{}s have to be easy to
generate, and the set of non-equivalent injective \LFT{}s has to be
large enough to make an exhaustive search intractable.

Several studies were made on the invertibility of \LFT{}s~\cite{Tao1973,Tao1988,Zongduo1996,Zongduo1998,Zongduo1999,ITA:9218509},
and some attacks to the FAPKC systems were presented~\cite{BaoIgarashi1995,Zongduo1996,TaoBook}.  However, as far
as we know, no study was conducted to determine the size of the key
space of these systems. 

Amorim \textit{et al} \cite{AMRciaa14} introduced a notion of
canonical \LFT{} and proved that each equivalence class has exactly
one canonical \LFT{}. Using this and a way to test if two \LFT{}s are
equivalent, they proved a result that allows to compute the size of
the equivalence class of a given \LFT{}. Two necessary and
sufficient conditions for a \LFT{} to be injective with some delay
$\tau$ are also well known \cite[Theorem 3.4]{ITA:9218509}. In this
paper we use these results to estimate the number of non-equivalent
\LFT{}s that are injective with some delay. The obtained estimate can
be used to compute the size of the key spaces of the mentioned
cryptographic systems. We also give a recurrence relation to count the
number of canonical \LFT{}s, and get an approximated value for the
percentage of equivalence classes formed by injective \LFT{}s. Knowing
this percentage is crucial to conclude if random generation of \LFT{}s
is a feasible option to generate keys. Several algorithms and
experimental results are also presented. All the algorithms were
implemented in \texttt{Python} using some 
\texttt{Sage}~\cite{sage} modules, to deal with matrices.

The paper is organized as follows. In Section~\ref{preliminaries} we
introduce the basic definitions and some preliminary results. In
Section~\ref{EstEquiClass} we start by presenting two algorithms, one
to test if a \LFT{} is injective with some delay $\tau$, and the other
to determine the equivalence class size of a given LFT{}. We then explain how these algorithms can be used to get an
approximate value for the number of non-equivalent \LFT{}s that are
injective with some delay. The recurrence relation to count the number
of canonical \LFT{}s is given in Section~\ref{EstProb}, as well as a
way to estimate the percentage of equivalence classes with injective
\LFT{}s. In Section~\ref{ExpRes} we present and discuss some
experimental results obtained using those algorithms.

\section{Preliminaries}\label{preliminaries}

As usual, for a finite set $A$, we let $|A|$ denote the cardinality of $A$, $A^n$ be the set of words of $A$ with length $n$, where $n\in\NN$, and $A^0=\{\varepsilon\}$, where
$\varepsilon$ denotes the empty word. We put $A^\star=\cup_{n\ge 0} A^n$,
the set of all finite words, and $A^\omega = \{a_0a_1 \cdots a_n
\cdots \mid a_i \in A \}$ is the set of infinite words.  Finally,
$|\alpha|$ denotes the length of $\alpha\in A^\star$.

The formal definition of a finite transducer (\FT{}) is the following.
\begin{definition}
  A \emph{finite transducer} is a quintuple
  $\left<\mathcal{X},\mathcal{Y},S,\delta,\lambda\right>$, where:
  $\mathcal{X}$ is a nonempty finite set, called the \emph{input
      alphabet};
   $\mathcal{Y}$ is a nonempty finite set, called the
    \emph{output alphabet};
   $S$ is a nonempty finite set called the \emph{set of states};
   $\delta: S\times \mathcal{X}\to S$, called the \emph{state
      transition function}; and
   $\lambda: S\times \mathcal{X}\to \mathcal{Y}$, called the
    \emph{output function}.
\end{definition}

Let $\trans$ be a finite transducer. The state transition function
$\delta$ and the output function $\lambda$ can be extended to finite
words, i.e.~elements of $\mathcal{X}^\star$, recursively, as follows:
\begin{align*}
  \delta(s,\varepsilon)&=s & \delta(s,x\alpha)&=\delta(\delta(s,x),\alpha)\\
  \lambda(s,\varepsilon)&=\varepsilon & 
  \lambda(s,x\alpha)&=\lambda(s,x)\>
  \lambda(\delta(s,x),\alpha), 
\end{align*}
where $s\in S$, $x\in\mathcal{X}$, and $\alpha \in \mathcal{X}^\star$.  In
an analogous way, $\lambda$ may be extended to $\mathcal{X}^\omega$.

From these definitions it follows that, for all $s\in S,
\alpha\in \mathcal{X}^\star$, and for all $\beta\in \mathcal{X}^\star\cup
\mathcal{X}^\omega$,
\begin{equation*}
  \lambda(s,\alpha\beta)=\lambda(s,\alpha)\>\lambda(\delta(s,\alpha),\beta).
\end{equation*}
A crucial concept to recall here is the concept of injective \FT. In
fact, there are two notions of injectivity that are behind the
invertibility property of \FT{}s used for cryptographic purposes: the
concept of $\omega$-injectivity and the concept of injectivity with
some delay $\tau$, with $\tau \in \NN$.

\begin{definition}
  A finite transducer $\trans$ is said to be \emph{$\omega$-injective}, if
  \begin{equation*}
    \forall s \in S, \> \forall \alpha, \alpha' \in \mathcal{X}^\omega, \quad
    \lambda(s,\alpha)=\lambda(s,\alpha') \implies \alpha=\alpha'. 
  \end{equation*}
  That is, for any $s\in S$, and any $\alpha \in \mathcal{X}^\omega$,
  $\alpha$ is uniquely determined by $s$ and $\lambda(s,\alpha)$.
\end{definition}
\begin{definition}
  A finite transducer $\trans$ is said to be \emph{injective with
    delay $\tau$} or \emph{$\tau$-injective}, with $\tau \in \NN$, if
  \begin{equation*}
    \forall s \in S, \> \forall x,x' \in \mathcal{X}, \> \forall
    \alpha, \alpha ' \in \mathcal{X}^\tau, \quad
    \lambda(s,x\alpha)=\lambda(s,x' \alpha') \implies x=x'. 
  \end{equation*}
  That is, for any $s \in S$, $x\in\mathcal{X}$, and $\alpha\in
  \mathcal{X}^\tau$, $x$ is uniquely determined by $s$ and
  $\lambda(s,x\alpha)$.
\end{definition}

It is quite obvious that if an FT{} is injective with some delay
$\tau \in \NN$, then it is injective with delay $\tau'$, for
$\tau'\geq \tau$, which implies that is also $\omega$-injective. The
reverse is also true. Tao~\cite[Corollary 1.4.3]{TaoBook} showed that
if $\trans$ is a $\omega$-injective \FT, then there exists a
non-negative integer $\tau\leq\frac{|S|(|S|-1)}{2}$ such that $M$ is
$\tau$-injective.

The notions of equivalent states and minimal transducer considered
here are the classical ones.

\begin{definition}
  Let
  $M_1=\left<\mathcal{X},\mathcal{Y}_1,S_1,\delta_1,\lambda_1\right>$
  and
  $M_2=\left<\mathcal{X},\mathcal{Y}_2,S_2,\delta_2,\lambda_2\right>$
  be two \FT{}s. Let $s_1 \in S_1$, and $s_2 \in S_2$. One says that
  $s_1$ and $s_2$ are \emph{equivalent}, and denotes this relation by
  $s_1 \sim s_2$, if $\forall \alpha \in \mathcal{X}^\star,\> \lambda_1(s_1,\alpha)=\lambda_2(s_2,\alpha)$.
\end{definition}
\begin{definition}
  A finite transducer $\trans$ is called \emph{minimal} if it has no
  pair of equivalent states.
\end{definition}
We now introduce the notion of equivalent transducers used in this
context.
\begin{definition}
  Let
  $M_1=\left<\mathcal{X},\mathcal{Y}_1,S_1,\delta_1,\lambda_1\right>$
  and
  $M_2=\left<\mathcal{X},\mathcal{Y}_2,S_2,\delta_2,\lambda_2\right>$
  be two \FT{}s. $M_1$ and $M_2$ are said to be \emph{equivalent}, and
  we denote this by $M_1 \sim M_2$, if the following two conditions are
  satisfied:
   $\forall s_1 \in S_1, \> \exists s_2 \in S_2: \> s_1 \sim
   s_2\quad \text{and}\quad \forall s_2 \in S_2, \> \exists s_1 \in
   S_1: \> s_1 \sim s_2.$
\end{definition}
This relation $\sim$ is an equivalence relation on the set of
\FT{}s. To simplify, an equivalence class formed by
$\omega$-injective \FT{}s is said to be
$\omega$-injective. Analogously, an equivalence class of
$\tau$-injective \FT{}s, for some $\tau \in \NN$, is said to be
$\tau$-injective.

Finally, we give the definition of what is called a linear finite transducer (\LFT{}).

\begin{definition}
  If $\mathcal{X}, \mathcal{Y}$ and $S$ are vector spaces over a field
  $\FF$, and both $\delta: S\times \mathcal{X} \to S$ and $\lambda: S
  \times \mathcal{X} \to \mathcal{Y}$ are linear maps, then the finite
  transducer $\trans$ is called \emph{linear} over $\FF$, and
  we say that $dim(S)$ is the size of $M$.
\end{definition}
If $\mathcal{X}, \mathcal{Y}$, and $S$ have dimensions $\ell$, $m$ and
$n$, respectively, then there exist matrices $A \in
\mathcal{M}_{n,n}(\FF)$, $B\in \mathcal{M}_{n,\ell}(\FF)$, $C \in
\mathcal{M}_{m,n}(\FF)$, and $D \in \mathcal{M}_{m,\ell}(\FF)$, such that
\begin{equation*}
  \delta(s,x) = A s+B x\> \text{ and }\> \lambda(s,x) = C s+D x,  
\end{equation*}
for all $s \in S, x \in \mathcal{X}$.  The matrices $A, B, C, D$ are
called the \LFT{} \emph{structural matrices}, and $\ell, m, n$ are called
the \LFT{} \emph{structural parameters}. An \LFT{} such that $C$ is the
null matrix (with the adequate dimensions) is called trivial.

Let $\mathcal{L}$ be the set of \LFT{}s over a field $\FF$, and let
$\mathcal{L}_n$ denote the set of \LFT{}s of size $n$. The
restriction of the equivalence relation $\sim$ to $\mathcal{L}$ is
also represented by $\sim$. Its restriction to $\mathcal{L}_n$ is
denoted by $\sim_n$.  

\begin{definition}
Let $M\in \mathcal{L}_n$ with structural matrices $A,B,C,D$. The
matrix
  \begin{equation*}
    \Delta_M=\left[
      \begin{array}{l}
        C\\
        CA\\
        \hphantom{C}\vdots\\
        CA^{n-1}
      \end{array}\right]
  \end{equation*}
  is called the \emph{diagnostic matrix} of $M$. 
\end{definition}
Amorim \textit{et al}~\cite{AMRciaa14} introduced a notion of
canonical \LFT{} and proved that each equivalence class of \LFT{}s has
exactly one canonical \LFT{}, which is minimal. Here, we recall the notions of canonical \LFT{} and of standard basis, used to define it.

\begin{definition}
Let $V$ be a $k$-dimensional vector subspace of
$\FF^n$, where $\FF$ is a field. The unique basis
$\{b_1,b_2,\ldots,b_k\}$ of $V$ such that the matrix $[b_1
\; b_2\; \cdots \;b_k]^T$ is in row echelon form will be here referred
to as the \emph{standard basis} of $V$.
\end{definition}

\begin{definition}
  Let $M=\left<\mathcal{X},\mathcal{Y},S,\delta,\lambda\right>$ be a
  linear finite transducer. One says that $M$ is a \emph{canonical
    \LFT{}} if $\{\Delta_Me_1,\Delta_Me_2,\cdots,\Delta_Me_n\}$ is the standard
  basis of $\{\Delta_Ms \mid s\in S\}$, where $\{e_1,e_2,\cdots,e_n\}$
  is the standard basis of $S$.
\end{definition}
In the same work it is also proved a fundamental result
about the size of \LFT{}s equivalence classes. It gives a way to
compute the number of \LFT{}s in $\mathcal{L}_{n_2}$ that are
equivalent to minimal \LFT{}s in $\mathcal{L}_{n_1}$, for $n_2\geq
n_1$. This result will be essential in Section~\ref{EstProb} to deduce
the recurrence relation that gives the number of canonical \LFT{}s.
\begin{theorem}\label{mainteo}
  Let $M_1$ be a minimal \LFT{} over $\FF_q$ with structural parameters $\ell,m, n_1$, and let $n_2 \geq n_1$. Then, the number of
  finite transducers $M \in\mathcal{L}_{n_2}$ which are equivalent to
  $M_1$ is $ (q^{n_2}-1) (q^{n_2}-q) \cdots (q^{n_2}-q^{r-1})
  q^{(n_2+\ell)(n_2-r)}$, where $r=\rank(\Delta_{M_1})$.
\end{theorem}
 We now recall
the notion of Smith normal form ($\SNF$) of a matrix and the well
known result (see \cite{jacobson85:_basic_algeb_i} or \cite[Theorem
II.9]{Newman1972}) that ensures its existence.
\begin{theorem}\label{SNF} Let $R$ be a principal ideal
  domain. Every matrix $A\in \mathcal{M}_{m,n}(R)$ is equivalent to a
  matrix of the form
  {\small\begin{equation*}
      \mathcal{D}=diag(d_1,d_2,\dots,d_r,0,\dots,0)=
      \left[
        \begin{array}{ccccccc}
          d_1&&&&&&\\
          &\ddots&&&\multicolumn{2}{c}{\multirow{1}*{{\Large \emph{0}}}}\\
          &&d_r&&&&\\
          &&&0&&&\\
          \multicolumn{3}{c}{\multirow{1}*{{\Large \emph{0}}}}&&\ddots&&\\
          &&&&&0&
      \end{array}
    \right]
  \end{equation*}}where $r=\rank(A)$, $d_i \not=0$ and $d_i \mid d_{i+1}$, \textit{i.e.}~$d_i$ divides $d_{i+1}$, for
$1\leq i \leq r-1$. The matrix $\mathcal{D}$ is called the \emph{Smith normal
  form} of $A$, and the elements $d_i$ are called the invariant
factors of $A$.
\end{theorem}
\section{Estimation of the number of $\tau$-injective equivalence
  classes}\label{EstEquiClass}
In this section we show how to estimate the number of non-equivalent
\LFT{}s that are $\tau$-injective, for some $\tau \in \NN$, by
generating \LFT{}s at random. Subsection~\ref{checkinvertibility} is
devoted to explain how to implement an algorithm in \texttt{Python} to test
if a given \LFT{} is injective with some delay $\tau$ using the
\texttt{Sage} system. In Subsection~\ref{sec:sizeEC} we present an
algorithm that, given a \LFT, computes the size of its equivalence
class.  Finally, in Subsection~\ref{sec:estimation} we explain how
these algorithms can be used to get an approximated value for the
number of $\tau$-injective equivalence classes, {\it i.e.}, the number
of non-equivalent \LFT{}s.

\subsection{Checking if a \LFT{} is injective with delay
  $\tau \in \NN$} \label{checkinvertibility}
Let $\trans$ be a \LFT{} over a field $\FF$ defined by the structural matrices $A$, $B$, $C$, $D$ and with structural parameters $\ell,m,n$. Starting at a state $s_0$ and reading an input sequence
$x_0x_1x_2\ldots$, one gets a sequence of states $s_0s_1s_2\ldots$ and
a sequence of outputs $y_0y_1y_2\ldots$ satisfying the relations $s_{t+1}=\delta(s_t,x_t)=A s_t+B x_t$ and $y_{t}=\lambda(s_t,x_t)=Cs_t+D x_t$, for all $t\geq 0$. Now, let $$X(z)=\sum_{t\geq0}x_tz^t,\quad Y(z)=\sum_{t\geq0}y_tz^t,\quad
Q(z)=\sum_{t\geq0}s_tz^t,$$ regarded as elements of the
$\FF[[z]]$-modules $\FF[[z]]^\ell$, $\FF[[z]]^m$, $\FF[[z]]^n$,
respectively, where $\FF[[z]]$ is the ring of formal power series over
$\FF$. Amorim \textit{et al}~\cite{ITA:9218509} showed that
\begin{equation*}
  Y(z)=G(z)s_0+H(z)X(z)
\end{equation*}
where $G(z)=C(I-Az)^{-1}$ and $H(z)=C(I-Az)^{-1}Bz+D$. The matrices $G\in\mathcal{M}_{m,n}(\FF)[[z]]$ and $H\in\mathcal{M}_{m,\ell}(\FF)[[z]]$ are called, respectively, the
\emph{free response matrix} and the \emph{transfer function matrix} of
the transducer. In the same paper the authors also proved that
\begin{equation}\label{defH}
H(z) =
\frac{1}{f(z)}
    \left(C(I-Az)^\star Bz+f(z)D\right),
\end{equation}
where $f(z)=\det(I-Az)$, and $P^\star$ denotes the adjoint matrix of
$P$. Consider the multiplicatively closed set
$\mathcal{S}=\left\{1+zb(z) \mid b(z) \in \FF[z] \right\}$, and let
$\FF[z]_{\mathcal{S}}=\left\{\frac{f}{g} \mid f \in \FF[z], g\in
  \mathcal{S}\right\}$ be the ring of fractions of $\FF[z]$ relative
to $\mathcal{S}$. Then, the transfer function matrix of a \LFT{} is in
$\mathcal{M}(\FF[z]_\mathcal{S})$. Since $\FF[z]_\mathcal{S}$ is a
principal ideal domain, and $z$ is its unique irreducible element, up
to units, the SNF of every transfer function matrix $H(z)$, with rank
$r$, is of the form
\begin{equation*}
  \mathcal{D}_{n_0,n_1,\dots,n_u}=
  diag(I_{n_0},zI_{n_1},\dots,z^uI_{n_u},0,\dots,0) =
  {\small
    \left[
        \begin{array}{cccccccc}
          I_{n_0}\\
          &&zI_{n_1}&&&\multicolumn{2}{c}{\multirow{1}*{{\large \emph{0}}}}\\
          &&&\ddots\\
          &&&&z^uI_{n_u}\\
          &&&&&0\\
           &\multicolumn{2}{c}{\multirow{1}*{{\large \emph{0}}}}&&&&\ddots\\
           &&&&&&&0
      \end{array}
    \right]},
\end{equation*}
where $n_i\geq0$, for $0\leq i \leq u$; $n_u \not =0$ unless $H(z)=0$,
and $\sum_{i = 0}^{u}n_i=r$.  We now restate the result~\cite{ITA:9218509} that gives two
necessary and sufficient conditions for a transducer to be injective
with some delay $\tau$. We put $n_i=0, \forall i>u$.
\begin{theorem}\label{teo6}
  Let $\mathcal{X}, \mathcal{Y}$ and $S$ be vector spaces over a field
  $\FF$, with dimensions $\ell$, $m$ and $n$, respectively. Let $\trans$
  be a \LFT{}, and let $H \in
  \mathcal{M}_{m,\ell}(\FF[z]_\mathcal{S})$ be its transfer function
  matrix. Let $\mathcal{D}=\mathcal{D}_{n_0,n_1,\dots,n_u}$ be the
  Smith normal form of $ H$, and assume $n_u \not=0$. Then, the
  following conditions are equivalent:
  \begin{enumerate}[(i)]
  \item $M$ is injective with delay $\tau$ \label{item1};\label{point1}
  \item $\sum_{i=0}^{\tau}n_i=\ell$ \label{item2};\label{point2}
  \item there is $H'\in \mathcal{M}_{\ell,m}(\FF[z]_\mathcal{S})$ such that
    $H'H=z^{\tau}I$.\label{item3}
  \end{enumerate}
\end{theorem}
In Algorithm~\ref{fig:isinvertible} one can read the definition of the function
\texttt{IsInjective}, which tests if a \LFT{} over $\FF_2$, defined by
its structural matrices, $A,B,C,D$, is $\tau$-injective by checking condition~(\ref{point2}) of the previous theorem.
\begin{lstlisting}[caption={Testing if a \LFT{} over $\FF_2$ is injective with some
    delay $\tau$.},label={fig:isinvertible}]
  def $\text{IsInjective}(A,B,C,D,tau)$: 
      $Ring = GF(Integer(2))['z']$
      $(z,) = Ring.\_first\_ngens(1)$
      $poly = \text{identity\_matrix}(A.\text{nrows}())-A*z$
      $fH = C*poly.\text{adjoint}()*B*z+poly.\text{det}()*D$
      $D\_fH = fH.\text{elementary\_divisors}()$
      $D\_H = [i.\text{gcd}(z**(tau+1))$ for $i$ in $D\_fH$ if $i \not= 0 ]$
      return $B.\text{ncols}() == \text{len}([j $ for $ j$ in $D\_H$ if $j <= z**tau])$
\end{lstlisting}
In lines 4--7 the $\SNF$ of $H(z)$ is computed as follows. It starts
by using the \texttt{Sage} function \texttt{elementary\_divisors} to determine the
invariant factors of $f(z)H(z)\in\mathcal{M}_{m,\ell}(\FF_2[z])$. Since
units are irrelevant in the $\SNF$ computation, and $f(z)$ is a unit
in $\FF_2[z]_\mathcal{S}$, one can find the invariant factors of $H(z)$
from the invariant factors of the matrix $f(z)H(z)$ using the
following straightforward result.
\begin{proposition}\label{simpSNF}
  Let $\mathcal{D}_{fH}=diag(d'_1,d'_2,\dots,d'_r,0,\dots,0) $ be the
  $\SNF$ of $f(z)H(z)$ and
  $\mathcal{D}_H=diag(d_1,d_2,\dots,d_r,0,\dots,0)$ the $\SNF$ of
  $H$. Then,
  \begin{equation}\label{eq:simpSNF}
    \forall i \in \{1,\dots,r\}, \> d_i=\gcd(d'_i,z^{u}),
  \end{equation} 
  where $z^{u}$ is the biggest power of $z$ that divides
  $d'_r$.
\end{proposition}
Having this, and since the entries of the matrix $f(z)H(z)$ belong to $\FF_2[z]$, the algorithm starts by defining the ring $\FF_2[z]$ (line 2), and $z$ as a variable in that ring (line 3).  
The expression \texttt{identity\_matrix(A.nrows())}, as the name suggests, returns the identity matrix whose size is the number of rows of $A$,
{\it i.e.}, $n$. 
The matrix $f(z)H(z)$ is then computed using the expression~\eqref{defH}, and the algorithm uses functions \texttt{adjoint} and \texttt{det}, to compute the adjoint and the determinant of a matrix, respectively (line 5). 
The invariant factors of $f(z)H(z)$ are computed using the
function \texttt{elementary\_divisors} (line 6). Since to check if
condition~\eqref{point2} of Theorem~\ref{teo6} is verified one just
needs to count the invariant factors of $H(z)$ that are less or equal
to $z^\tau$, we apply Proposition~\ref{simpSNF} in the algorithm by
replacing $z^u$ with $z^{\tau+1}$ in expression~\eqref{eq:simpSNF}
(line 7). The algorithm then returns \texttt{True} if the number of
invariant factors of $H(z)$ which divide $z^\tau$ is equal to $\ell$,
\textit{i.e.}, is equal to the number of columns of the matrix $B$. It
returns \texttt{False} otherwise.
\subsection{Determining the size of equivalence
  classes}\label{sec:sizeEC}
In a previous work~\cite{AMRciaa14}, it was proved that, given a \LFT{} over $\FF_q$, $M$, with structural matrices $A,B,C,D$ and structural parameters
$\ell,m,n \in \NN \setminus \{0\}$, the size of $[M]_{\sim_n}$ is given by the following expression:
\begin{equation}\label{eq:eqsize}
  \left|[M]_{\sim_n}\right|=\prod_{i=0}^{r-1}\left(q^{n}-q^i\right) \cdot q^{\left(n+\ell\right)\left(n-r\right)}, \text{ where } r=\rank\left(\Delta_M \right).
\end{equation}
Algorithm~\ref{fig:sizeclass} shows the definition of
\texttt{EquivClassSize} that computes the size of an equivalence class
using expression~\eqref{eq:eqsize} for $q=2$. This takes as input the
structural matrices $A,B,C,D$, and the structural parameters $\ell,m,n$
are determined using functions \texttt{nrows} and
\texttt{ncols} (lines 2--4). To determine the value of $r$, it calls functions $\texttt{stack}$ and $\texttt{rank}$. The
first is used to create the \LFT{} diagnostic matrix (lines 5--7), and
the second is used to determine the rank of that matrix (line 8). The
size of the equivalence class is then easily obtained through a loop
(lines 9--12).
\begin{lstlisting}[caption={Determining the size of equivalence
    classes.},label={fig:sizeclass}]
  def $\text{EquivClassSize}(A,B,C,D)$:
      $l = B.\text{ncols}()$
      $m = C.\text{nrows}()$ 
      $n = A.\text{nrows}()$
      $K = \text{copy.deepcopy}(C)$
      for $j$ in $\{1,\dots,n-1\}$:
          $K = K.\text{stack}(K*A)$
      $r = K.\text{rank}()$
      $size = 1$
      for $j$ in $\{0,\dots,r-1\}$:
          $size = size*(2**n-2**j)$
      $size = size*2**((n+l)*(n-r))$
      return $size$
\end{lstlisting}
%
\subsection{Computing an approximated value for the number of
  $\tau$-injective equivalence classes}\label{sec:estimation}

Let $\mathcal{E}$ be the set of equivalence classes of \LFT{}s over $\FF_q$ with
structural parameters $\ell,m,n$. Let
$\mathcal{I_\tau}\subseteq\mathcal{E}$ be the set of the
$\tau$-injective equivalence classes, \textit{i.e.}, $\mathcal{I_\tau}
=\left\{[M]_{\sim_n} \in \mathcal{E} \mid M \text{ is } \tau
  \text{-injective}\right\} $. One wants to estimate
$|\mathcal{I_\tau}|$.

It is easy to generate random \LFT{}s because, given a triple of structural
parameters, one just needs to generate structural matrices $A,B,C,D$
with the appropriate sizes. Using the method described in
Subsection~\ref{checkinvertibility}, it is also possible to test if a
\LFT{} is injective with some delay $\tau$. Hence, one can get an approximated value for
$|\mathcal{I_\tau}|$ with simple
random sampling, as we will see in the remaining of this
subsection.

Let $\mathcal{L}_{\ell,m,n}$ be the set of \LFT{}s with structural
parameters $\ell,m,n$. Let
$\mathcal{R}$ be a multiset of randomly generated \LFT{}s in
$\mathcal{L}_{\ell,m,n}$, and $\eta_E$ the number of occurrences in
$\mathcal{R}$ of transducers that belong to a class
$E\in\mathcal{E}$. Let $p_E$ be the probability that a \LFT{} in $\mathcal{L}_{\ell,m,n}$
is in the class $E \in \mathcal{E}$, that is, $p_E =\frac
{|E|}{|\mathcal{L}_{\ell,m,n}|}$. One knows that $\frac{\eta_E}{|\mathcal{R}|}$
is an approximated value for $p_E$, and that the larger the sample
size $|\mathcal{R}|$, the better will the approximation be.

Take $E\in\mathcal{E}$, and let:
\begin{equation}
  \mu_E = \left\{ \begin{array}{ll} \frac{1}{p_E} & \text{if } E \in \mathcal{I}_\tau\\ 0 & \text{otherwise} \end{array} \right..
\end{equation}
Trivially,
\begin{equation*}
  |\mathcal{\mathcal{I}_\tau}| = \sum_{E\in \mathcal{I}_\tau}1= \sum_{E\in \mathcal{I}_\tau}p_E\frac{1}{p_E}=\sum_{E\in \mathcal{E}}p_E\mu_E.
\end{equation*}
Consequently,
\begin{equation*}
  |\mathcal{I_\tau}| \approx \sum_{E \in \mathcal{E}} \frac{\eta_E}{|\mathcal{R}|}\mu_E = \frac{1}{|\mathcal{R}|} \sum_{E \in \mathcal{E}} \eta_E\mu_E.
\end{equation*}
Since, obviously,  $M \in E$ if and only if $E=[M]_{\sim_n}$, one has
\begin{equation*}
\sum_{E \in \mathcal{E}} \eta_E\mu_E = \sum_{M\in \mathcal{R}}\mu_{[M]_\sim}, \text{ and } |\mathcal{I_\tau}| \approx \frac{1}{|\mathcal{R}|}\sum_{M\in \mathcal{R}}\mu_{[M]_\sim}.
\end{equation*}
Therefore we can get an approximated value for $|\mathcal{I_\tau}|$
using a simple function as the one presented in
Algorithm~\ref{fig:estimativa}.
\begin{lstlisting}[caption={Estimating the number of non-equivalent
    \LFT{}s.},label={fig:estimativa}]
  def $EstCountInjective(nr,l,m,n,tau)$: 
      count = 0
      for $i$ in $\{1,\dots,nr\}$:
          $A,B,C,D = \text{RandomLFT}(l,m,n)$
          if $\text{IsInjective}(A,B,C,D,tau)$:
             $count = count + 1/\text{Probability}(A,B,C,D)$
      return $count/nr$
\end{lstlisting}
The function \texttt{EstCountInjective} takes as input the sample size,
represented by the variable \texttt{nr}, the structural parameters
$\ell,m,n$, and the delay $\tau$. It calls the following three functions:
\begin{itemize}
\item \texttt{RandomLFT}: a function such that, given the parameters
  $\ell,m,$ and $n$, returns matrices $A \in
  \mathcal{M}_{n,n}(\FF_2)$, $B\in \mathcal{M}_{n,\ell}(\FF_2)$, $C \in
  \mathcal{M}_{m,n}(\FF_2)$, and $D \in \mathcal{M}_{m,\ell}(\FF_2)$,
  whose entries were uniformly randomly generated;
\item \texttt{IsInjective}: the function defined in
  Subsection~\ref{checkinvertibility};
\item \texttt{Probability}: a function such that, given the
  structural matrices of a \LFT{}, $M$, it returns $p_{[M]_ \sim}$
  using the function \texttt{EquivClassSize} presented in
  Subsection~\ref{sec:sizeEC}.
\end{itemize}
Given an input, the algorithm starts by initializing the variable
\texttt{count} with the value $0$ (line 2). Then, at each iteration of
the loop, generates a \LFT{}, let us say $M$, and if $M$ is injective
with delay $\tau$ it adds, to the variable \texttt{count}, the value
of $\mu_{[M]_\sim}$ (lines 3--6). This way, when the loop is finished,
one has $\texttt{count}= \sum_{M\in \mathcal{R}}\mu_{[M]_\sim},$ where
$\mathcal{R}$ is the set of the \texttt{nr} random generated
\LFT{}s. It returns $\texttt{count} / \texttt{nr}$, that is, an estimate for
$|\mathcal{I_\tau}|$.

\section{Estimating the percentage of $\tau$-injective equivalent
  classes}\label{EstProb}
 
In this section, we first deduce a recurrence relation that, given
$\ell,m,n \in \NN\setminus \{0\}$, counts the number of canonical \LFT{}s over $\FF_q$
with structural parameters $\ell,m,n$. Then we show how to estimate the
percentage of $\tau$-injective equivalence classes.

Let $\ell,m,n \in \NN\setminus \{0\}$, and consider the following
notation:
\begin{itemize}
\item $\LT_{\ell,m,n}$ denotes the total number of \LFT{}s over $\FF_q$ in $\mathcal{L}_{\ell,m,n}$;
\item $\TT_{\ell,m,n}$ denotes the number of trivial \LFT{}s over $\FF_q$ in $\mathcal{L}_{\ell,m,n}$;
\item $\TM_{\ell,m,n}$ denotes the number of non-trivial \LFT{}s over $\FF_q$ in
  $\mathcal{L}_{\ell,m,n}$ that are minimal;
\item $\TNM_{\ell,m,n}$ denotes the number of non-trivial \LFT{}s over $\FF_q$
  in $\mathcal{L}_{\ell,m,n}$ that are not minimal;
\item $\CT_{\ell,m,n}$ denotes the number of canonical \LFT{}s over $\FF_q$ in
  $\mathcal{L}_{\ell,m,n}$.
\end{itemize}
It is obvious that $\LT(\ell,m,n) = \TT(\ell,m,n) + \TNM(\ell,m,n) + \TM(\ell,m,n)$.

The number of trivial transducers is easy to find: since a \LFT{} is
trivial when $C=0$, the entries of the other matrices ($A,B,$ and $D$)
can take any value and, therefore,
\begin{equation*}
\TT(\ell,m,n)=q^{n^2+\ell(m+n)}.
\end{equation*}
The set of non-trivial \LFT{}s in $\mathcal{L}_{\ell,m,n}$ that are minimal is
formed by the equivalence classes that have a canonical \LFT{}. Notice
that, from Theorem~\ref{mainteo}, those classes all have the same
cardinality. Let $\EC(n)$ be the size of the equivalence class
$[M]_{\sim_n}$, where $M$ is a canonical transducer in
$\mathcal{L}_{\ell,m,n}$. Then, also from Theorem~\ref{mainteo}, $\EC(n) =
\prod_{i=0}^{n-1}(q^n-q^i)$. Therefore,
\begin{equation*}
  \TM(\ell,m,n) = \EC(n)\cdot\CT(\ell,m,n)=\prod_{i=0}^{n-1}(q^n-q^i)\cdot\CT(\ell,m,n).
\end{equation*}
Now, let us see how to determine $\TNM(\ell,m,n)$ for all
$\ell,m,n \in \NN\setminus\{0\}$. 

For $n=1$, all the non-trivial
\LFT{}s are canonical. Therefore $\TNM(\ell,m,1) = 0$, and
\begin{equation}\label{CT1}
  \CT(\ell,m,1)  = \LT(\ell,m,1)-\TT(\ell,m,1)
\end{equation}
For $n=2$, $\TNM(\ell,m,n)$ is the number of transducers in $\LL_{\ell,m,2}$ that
are equivalent to transducers in $\LL_{\ell,m,1}$. Theorem~\ref{mainteo} tells
us a way to compute the number of \LFT{}s in $\LL_{\ell,m,n_2}$ that are
equivalent to minimal transducers in $\LL_{\ell,m,n_1}$, for $n_2\geq
n_1$. Let $\NM(\ell,n_1,n_2)$ be that value, that is, $\NM(\ell,n_1,n_2) =
\prod_{i=0}^{n_1-1}(q^{n_2}-q^i) \cdot q^{\left(n_2+\ell\right)
  \left(n_2-n_1 \right)}$. Then,
\begin{equation*}
  \TNM(\ell,m,2) = \CT(\ell,m,1)\cdot \NM(\ell,1,2)
  = \CT(\ell,m,1)\cdot (q^{2}-1) \cdot q^{\ell+2}
\end{equation*}
For $n=3$, the set of non-minimal \LFT{}s if formed by the \LFT{}s
that are equivalent to minimal transducers in $\LL_{\ell,m,1}$, and the ones
that are equivalent to minimal transducers in $\LL_{\ell,m,2}$. Therefore,
\begin{eqnarray*}
  \TNM(\ell,m,3) &=& \CT(\ell,m,1)\cdot \NM(\ell,1,3) + \CT(\ell,m,2)\cdot \NM(\ell,2,3) \\
  &=& \sum_{i=1}^{2}\CT(\ell,m,i) \cdot  \NM(\ell,i,3) = \sum_{i=1}^{2}\CT(\ell,m,i) \cdot \prod_{j=0}^{i-1}(q^{3}-q^j) \cdot q^{\left(\ell+3\right) \left(3-i \right)}
\end{eqnarray*}
This process can be generalized to get:
\begin{equation*}
  \TNM(\ell,m,n) =\sum_{i=1}^{n-1}\CT(\ell,m,i) \cdot
  \NM(\ell,i,n).
\end{equation*}
Therefore, given $\ell,m,n \in \NN\setminus\{0\}$, the number of
canonical \LFT{}s with structural parameters $\ell,m,n$ satisfies the
following recurrence relation:
\begin{equation*}
  \left\{
    \begin{array}{lcl}
      \CT(\ell,m,1)  &=& (q^{m}-1)q^{\ell(m+1)+1}\vspace{0.3cm}\\
      \CT(\ell,m,n)  &=&
      \frac{1}{\EC(n)}\cdot\left(\LT(\ell,m,n)-\TT(\ell,m,n)-\TNM(\ell,m,n)\right), \text{ for } n\geq2
    \end{array}
  \right.
\end{equation*}
where $$\LT(\ell,m,n) = q^{m\ell+n(\ell+m+n)},\>\>\> \EC(n) =
\prod_{i=0}^{n-1}(q^n-q^i), \>\>\> \TT(\ell,m,n) = q^{n^2+\ell(m+n)},$$ 
$$\TNM(\ell,m,n) = \sum_{i=1}^{n-1}{\CT(\ell,m,i) \cdot \NM(\ell,i,n) },\> \text{ and } \NM(\ell,i,n) = \prod_{j=0}^{i-1}(q^{n}-q^j) \cdot
q^{\left(n+\ell\right) \left(n-i \right)}.$$
We define \texttt{CountCT} (Algorithm~\ref{fig:countCT})
taking as input a triple $\ell,m,n \in \NN \setminus\{0\}$, and using
the previous recurrence relation to compute the number of canonical
\LFT{}s with structural parameters $\ell,m,n$. It starts by checking if
$n=1$ and, if that is true, it computes $\CT(\ell,m,1)$ using
expression~\eqref{CT1} (lines 2--3). If $n\geq 2$, it computes
$\EC(n)$, $\LT(\ell,m,n)$, $\TT(\ell,m,n)$ and $\TNM(\ell,m,n)$ using the
expressions given above.
\begin{lstlisting}[caption={Counting the number of canonical \LFT{}s.},label={fig:countCT}]
  def $CountCT(l,m,n)$: 
      if $n=1$:
         return $(2**m-1)*2**(l*(m+1)+1)$
      else:
         $EC = 1$
         for $i$ in $\{0,\dots,n-1\}$:
             $EC = EC *(2**n-2**i)$
         $LT = 2**(m*l+n*(l+m+n))$
         $TT = 2**(n**2+l*(m+n))$
         $TNM =0$ 
         for $i$ in $\{1,\dots,n-1\}$:
             $NM = 2**(n+l)*(n-i)$
             for $j$ in $\{0,\dots,i-1\}$:
                 $NM = NM*(2**n-2**j)$
             $TNM = TNM + \text{CountCT}(l,m,i)*NM$
      return $(LT-TT-TNM)/EC$
\end{lstlisting}
The function \texttt{CountCT}
computes the exact number of canonical \LFT{}s which have structural
parameters $\ell,m,n \in \NN\setminus\{0\}$. Thus, it can be used to
count the exact number of equivalence classes that contain at least
one \LFT{} with strucural parameters $\ell,m,n$. Given a triple
$\ell,m,n\in \NN\setminus\{0\}$, one just needs to sum up the number of
canonical \LFT{}s that have structural parameters $\ell,m,n',$ for
$n'\leq n$. Since the function \texttt{EstCountInjective} defined in
Algorithm~\ref{fig:estimativa} gives an approximate value for the
number of equivalence classes of \LFT{}s with structural parameters
$\ell,m,n \in \NN \setminus \{0\}$ that are $\tau$-injective, we can obtain, using these two functions, an estimated value
for the percentage of $\tau$-injective equivalence classes. The function
\texttt{EstPercInjective} (Algorithm~\ref{fig:estp}) implements this
process.
\begin{lstlisting}[caption={Estimating the percentage of $\tau$-injective equivalence classes.},label={fig:estp}]
  def $EstPercInjective(nr,l,m,n,tau)$: 
      $EC = 0$
      for $i$ in $\{1,\dots,n\}$: 
         $EC = EC + \text{CountCT}(l,m,i)$
      return $EstCountInjective(nr,l,m,n,tau)/EC$
\end{lstlisting}
\vspace{-0.5cm}\section{Experimental results}\label{ExpRes}
In this Section we present some experimental results on the number of $\omega$-injective and $\tau$-injective equivalent classes of \LFT{}s over $\FF_2$, for some values of $\tau \in \NN$. Recall that if a \LFT{} is $\tau$-injective for some $\tau \in \NN$, then it is $\omega$-injective.

For each triple of structural parameters $\ell,m,n$, with $\ell \in
\{1,\dots,5\}$, $m=5$ and $n\in\{1,\dots,10\}$, we uniformly randomly
generated a sample of $20\,000$ \LFT{}s. With these samples we estimate
the number of $\tau$-injective equivalence classes, for $\tau \in
\{0,1,\dots,10\}$, using \texttt{EstCountInjective}
defined above (Algorithm~\ref{fig:estimativa}). The total number of
equivalence classes was obtained using the recurrence relation
to count canonical \LFT{}s. Then, using the previous results, we
computed an approximated value for the percentage of $\tau$-injective
equivalence classes of \LFT{}s. The size of each sample is sufficient
to ensure the statistical significance with a 99$\%$ confidence level
within a 1$\%$ error margin. The sample size is calculated
with the formula $N = ( \frac{z}{2\epsilon})^2$, where $z$ is obtained
from the normal distribution table such that $P(-z < Z < z)) =
\gamma$, $\epsilon$ is the error margin, and $\gamma$ is the desired
confidence level.

In Table~\ref{tableIclasses}, we present the approximated values for the
number of $10$-injective equivalence classes when $m=5$, and $n, l$
range in $\{1,\dots,10\}$ and $\{1,\dots,5\}$, respectively. We
chose to show the results for $\tau=10$ because this value is large
enough to draw conclusions about the number of $\omega$-injective
equivalence classes. From the results obtained, one can observe an
exponential growth on the number of $10$-injective equivalence
classes, as $n$ and $\ell$ increase. Consequently, the number of
$\omega$-injective equivalence classes also grows exponentially.

\begin{table}[h]
  \centering\footnotesize
  \begin{tabular}{|c|c|c|c|c|c|c|}
    \cline{3-7}
    \multicolumn{1}{c}{}&\multicolumn{1}{c}{} & \multicolumn{5}{|c|}{\footnotesize $\ell$}\\ 
    \cline{3-7} 
    \multicolumn{2}{c|}{}&1&2&3&4&5\\\hline
    \multirow{10}{*}{\footnotesize $n$}&  1
&\scriptsize$3.91\times 10^{03}$&\scriptsize$2.42\times 10^{05}$&\scriptsize$1.44\times 10^{07}$&\scriptsize$7.66\times 10^{08}$&\scriptsize$2.97\times 10^{10}$\\
    \cline{2-7}
& 2 &\scriptsize$3.34\times 10^{05}$&\scriptsize$4.17\times 10^{07}$&\scriptsize$5.13\times 10^{09}$&\scriptsize$5.92\times 10^{11}$&\scriptsize$5.29\times 10^{13}$\\
    \cline{2-7}
& 3 &\scriptsize$2.45\times 10^{07}$&\scriptsize$6.15\times 10^{09}$&\scriptsize$1.54\times 10^{12}$&\scriptsize$3.70\times 10^{14}$&\scriptsize$7.39\times 10^{16}$\\
    \cline{2-7}
& 4 &\scriptsize$1.66\times 10^{09}$&\scriptsize$8.45\times 10^{11}$&\scriptsize$4.26\times 10^{14}$&\scriptsize$2.10\times 10^{17}$&\scriptsize$9.24\times 10^{19}$\\
    \cline{2-7}
& 5 &\scriptsize$1.10\times 10^{11}$&\scriptsize$1.12\times 10^{14}$&\scriptsize$1.13\times 10^{17}$&\scriptsize$1.14\times 10^{20}$&\scriptsize$1.05\times 10^{23}$\\
    \cline{2-7}
& 6 &\scriptsize$7.17\times 10^{12}$&\scriptsize$1.45\times 10^{16}$&\scriptsize$2.96\times 10^{19}$&\scriptsize$5.97\times 10^{22}$&\scriptsize$1.15\times 10^{26}$\\
    \cline{2-7}
& 7 &\scriptsize$4.61\times 10^{14}$&\scriptsize$1.87\times 10^{18}$&\scriptsize$7.64\times 10^{21}$&\scriptsize$3.10\times 10^{25}$&\scriptsize$1.22\times 10^{29}$\\
    \cline{2-7}
& 8 &\scriptsize$2.96\times 10^{16}$&\scriptsize$2.40\times 10^{20}$&\scriptsize$1.96\times 10^{24}$&\scriptsize$1.60\times 10^{28}$&\scriptsize$1.28\times 10^{32}$\\
    \cline{2-7}
&\ 9 &\scriptsize$1.90\times 10^{18}$&\scriptsize$3.08\times 10^{22}$&\scriptsize$5.04\times 10^{26}$&\scriptsize$8.24\times 10^{30}$&\scriptsize$1.33\times 10^{35}$\\
    \cline{2-7}
& 10 &\scriptsize$1.22\times 10^{20}$&\scriptsize$3.95\times 10^{24}$&\scriptsize$1.29\times 10^{29}$&\scriptsize$4.23\times 10^{33}$&\scriptsize$1.37\times 10^{38}$\\
    \hline
  \end{tabular}
  \caption{Approximated values for the number of injective classes when $m=5$ and $\tau=10$.}
  \label{tableIclasses}
\end{table}

The results on the percentage of $\tau$-injective equivalence classes
are exhibited in Tables~\ref{table25x}--\ref{table55x}. Each of these
tables presents the approximated percentage for a given value of $\ell \in
\{2,\dots,5\}$, while $n$ and $\tau$ range in $\{1,\dots,10\}$ and
$\{0,1,\dots,10\}$, respectively.

\begin{table}[h]
  \centering\scriptsize
  \begin{tabular}{|c|c|c|c|c|c|c|c|c|c|c|c|c|}
    \cline{3-13}
    \multicolumn{1}{c}{}&\multicolumn{1}{c}{} & \multicolumn{11}{|c|}{\footnotesize $\tau$} \\
    \cline{3-13} 
    \multicolumn{2}{c|}{}&\bf 0&\bf1&\bf2&\bf3&\bf4&\bf5&\bf6&\bf7&\bf8&\bf9&\bf10\\\hline
    \multirow{10}{*}{\footnotesize $n$}& \bf 1
    &90.88&95.21&95.21&95.21&95.21&95.21&95.21&95.21&95.21&95.21&95.21\\
    \cline{2-13}
    &\bf 2& 90.5&97.06&97.2&97.2&97.2&97.2&97.2&97.2&97.2&97.2& 97.2\\
    \cline{2-13}
    &\bf 3&90.82&98.27&98.58&98.62&98.62&98.62&98.62&98.62&98.62&98.62&98.62\\
    \cline{2-13}
    &\bf 4&91.1&99.07&99.53&99.57&99.57&99.57&99.57&99.57&99.57&99.57&99.57\\
    \cline{2-13}
    &\bf 5&91.01&99.18&99.72&99.74&99.74&99.74&99.74&99.74&99.74&99.74&99.74\\
    \cline{2-13}
    &\bf 6&91.07&99.37&99.92&99.95&99.96&99.96&99.96&99.96&99.96&99.96&99.96\\
    \cline{2-13}
    &\bf 7&90.75&99.12&99.69&99.73&99.73&99.73&99.73&99.73&99.73&99.73&99.73\\
    \cline{2-13}
    &\bf 8&90.64&99.31&99.76&99.81&99.81&99.81&99.81&99.81&99.81&99.81&99.81\\
    \cline{2-13}
    &\bf 9&90.6&99.18&99.7&99.74&99.74&99.75&99.75&99.75&99.75&99.75&99.75\\
    \cline{2-13}
    &\bf 10&90.85&99.39&99.85&99.89&99.89&99.89&99.89&99.89&99.89&99.89&99.89\\
    \hline
  \end{tabular}
  \caption{Approximated percentage value for $\ell=2$ and $m=5$.}
  \label{table25x}
\end{table}

In Table~\ref{table25x} we present the results for $\ell=2$. The results show
that, in this case, when $n$ increases, there is a significant
increase in the percentage of $\tau$-injective \LFT{}s, for $\tau \geq
1$. Nonetheless, when $n=1$ the percentage of $1$-injective (and
consequently $\omega$-injective) \LFT{}s is already very high (above
95$\%$). This suggests that, in this case, there is also a very high
probability of a uniform random generated \LFT{} be
$\omega$-injective.

The results for $\ell=3$, presented in Table~\ref{table35x}, also show a
significant growing of the values with $n$. A  more
careful observation of the column  $\tau =10$, allow us to conclude that
when  $n\geq 3=\ell$,  the  percentage of  $\omega$-injective \LFT{}s  is
above $95 \%$.
\begin{table}[h]
  \centering\scriptsize
  \begin{tabular}{|c|c|c|c|c|c|c|c|c|c|c|c|c|}
    \cline{3-13}
    \multicolumn{1}{c}{}&\multicolumn{1}{c}{} & \multicolumn{11}{|c|}{\footnotesize $\tau$}\\
    \cline{3-13} 
    \multicolumn{2}{c|}{}&\bf 0&\bf1&\bf2&\bf3&\bf4&\bf5&\bf6&\bf7&\bf8&\bf9&\bf10\\\hline
    \multirow{10}{*}{\footnotesize $n$}& \bf 1
    &79.42&88.48&88.48&88.48&88.48&88.48&88.48&88.48&88.48&88.48&88.48\\
    \cline{2-13}
    &\bf 2&79.08&92.77&93.61&93.61&93.61&93.61&93.61&93.61&93.61&93.61&93.61\\
    \cline{2-13}
    &\bf 3&79.19&94.98&96.54&96.68&96.68&96.68&96.68&96.68&96.68&96.68&96.68\\
    \cline{2-13}
    &\bf 4&79.22&96.31&98.27&98.47&98.48&98.48&98.48&98.48&98.48&98.48&98.48\\
    \cline{2-13}
    &\bf 5&79.69&96.89&99.04&99.28&99.29&99.29&99.29&99.29&99.29&99.29&99.29\\
    \cline{2-13}
    &\bf 6&79.68&97.14&99.39&99.66&99.70&99.71&99.71&99.71&99.71&99.71&99.71\\
    \cline{2-13}
    &\bf 7&79.21&97.37&99.58&99.79&99.83&99.85&99.85&99.85&99.85&99.85&99.85\\
    \cline{2-13}
    &\bf 8&79.72&97.22&99.52&99.79&99.82&99.82&99.82&99.82&99.82&99.82&99.82\\
    \cline{2-13}
    &\bf 9&79.50&97.32&99.56&99.85&99.90&99.91&99.91&99.91&99.91&99.91&99.91\\
    \cline{2-13}
    &\bf 10&80.07&97.64&99.83&100&100&100&100&100&100&100&100\\
    \hline
  \end{tabular}
  \caption{Approximated percentage value for $\ell=3$ and $m=5$.}
  \label{table35x}
\end{table}

In Tables~\ref{table45x} and \ref{table55x} we present the results for
$\ell=4$ and $\ell=5$, respectively. Again, the percentage of
$\omega$-injective \LFT{}s is quite high for $n \geq \ell$. 

\begin{table}[h]
  \centering\scriptsize
  \begin{tabular}{|c|c|c|c|c|c|c|c|c|c|c|c|c|}
    \cline{3-13}
    \multicolumn{1}{c}{}&\multicolumn{1}{c}{} & \multicolumn{11}{|c|}{\footnotesize $\tau$}\\
    \cline{3-13} 
    \multicolumn{2}{c|}{}&\bf 0&\bf1&\bf2&\bf3&\bf4&\bf5&\bf6&\bf7&\bf8&\bf9&\bf10\\\hline
    \multirow{10}{*}{\footnotesize $n$}& \bf 1
&59.09&73.64&73.64&73.64&73.64&73.64&73.64&73.64&73.64&73.64&73.63\\    \cline{2-13}
&\bf 2&59.70&81.83&84.60&84.60&84.60&84.60&84.60&84.60&84.60&84.60&84.60\\    \cline{2-13}
&\bf 3&59.50&85.53&90.49&91.07&91.07&91.07&91.07&91.07&91.07&91.07&91.07\\    \cline{2-13}
&\bf 4&59.76&87.83&93.95&95.01&95.13&95.13&95.13&95.13&95.13&95.13&95.13\\    \cline{2-13}
&\bf 5&59.01&88.77&95.79&97.35&97.60&97.64&97.64&97.64&97.64&97.64&97.64\\    \cline{2-13}
&\bf 6&59.58&89.29&96.39&98.14&98.48&98.52&98.53&98.53&98.53&98.53&98.53\\    \cline{2-13}
&\bf 7&59.93&89.49&96.97&98.76&99.14&99.19&99.22&99.22&99.22&99.22&99.22\\    \cline{2-13}
&\bf 8&59.43&89.30&97.14&98.87&99.35&99.49&99.51&99.51&99.51&99.51&99.51\\    \cline{2-13}
&\bf 9&59.93&89.91&97.40&99.31&99.81&99.95&99.97&99.98&99.98&99.98&99.98\\    \cline{2-13}
&\bf 10&59.81&89.46&97.64&99.51&99.99&100&100&100&100&100&100\\    
\hline
  \end{tabular}
  \caption{Approximated percentage value for $\ell=4$ and $m=5$.}
  \label{table45x}
\end{table}
\begin{table}[h]
  \centering\scriptsize
  \begin{tabular}{|c|c|c|c|c|c|c|c|c|c|c|c|c|}
    \cline{3-13}
    \multicolumn{1}{c}{}&\multicolumn{1}{c}{} & \multicolumn{11}{|c|}{\footnotesize $\tau$}\\
    \cline{3-13} 
    \multicolumn{2}{c|}{}&\bf 0&\bf1&\bf2&\bf3&\bf4&\bf5&\bf6&\bf7&\bf8&\bf9&\bf10\\\hline
    \multirow{10}{*}{\footnotesize $n$}& \bf 1
&29.29&44.63&44.63&44.63&44.63&44.63&44.63&44.63&44.63&44.63&44.63\\    \cline{2-13}
&\bf 2&30.26&53.48&59.11&59.11&59.11&59.11&59.11&59.11&59.11&59.11&59.11\\    \cline{2-13}
&\bf 3&29.75&57.69&68.60&71.09&71.09&71.09&71.09&71.09&71.09&71.09&71.09\\    \cline{2-13}
&\bf 4&30.13&61.15&75.19&80.37&81.63&81.63&81.63&81.63&81.63&81.63&81.63\\    \cline{2-13}
&\bf 5&29.96&62.07&78.05&84.84&87.21&87.74&87.74&87.74&87.74&87.74&87.74\\    \cline{2-13}
&\bf 6&29.21&62.69&79.92&88.01&91.37&92.52&92.79&92.79&92.79&92.79&92.79\\    \cline{2-13}
&\bf 7&29.35&62.63&80.43&88.92&92.98&94.87&95.50&95.65&95.65&95.65&95.65\\    \cline{2-13}
&\bf 8&29.78&63.60&81.02&90.20&94.50&96.43&97.33&97.62&97.67&97.67&97.67\\    \cline{2-13}
&\bf 9&30.07&63.39&81.08&90.05&94.57&96.71&97.85&98.35&98.46&98.50&98.50\\    \cline{2-13}
&\bf 10&28.97&62.58&80.92&90.70&95.22&97.24&98.34&98.87&99.14&99.25&99.26\\    
\hline
  \end{tabular}
  \caption{Approximated percentage value for $\ell=5$ and $m=5$.}
  \label{table55x}
\end{table}

Observing all the tables, it can be noticed that the approximated
percentage value, specially for low values of $n$, suffers a big
reduction when $\ell$ increases from $1$ to $5$. However, the
growth, as a function of $n$, is much steeper for higher values of
$\ell$. This ensures that, for a not so large value of $n$, the
percentage of $\omega$-injective \LFT{}s is very high. Therefore, if
one uniformly random generates \LFT{}s, it is highly probable to get $\omega$-injective ones.
 
We give here the results of an additional experiment, taking $\ell=m=8$, $n \in \{1,\dots,10\}$ and $\tau \in \{0,1,\dots,10\}$. The percentages of $\tau$-injective \LFT{}s obtained are presented in Table~\ref{table88x}. Again, for values of $n$ slightly larger than $\ell$ and $m$, one can see that  the percentage of $\omega$-injective \LFT{}s is very high. 
\begin{table}[h]
  \centering\scriptsize
  \begin{tabular}{|c|c|c|c|c|c|c|c|c|c|c|c|c|}
    \cline{3-13}
    \multicolumn{1}{c}{}&\multicolumn{1}{c}{} & \multicolumn{11}{|c|}{\footnotesize $\tau$}\\
    \cline{3-13} 
    \multicolumn{2}{c|}{}&\bf 0&\bf1&\bf2&\bf3&\bf4&\bf5&\bf6&\bf7&\bf8&\bf9&\bf10\\\hline
    \multirow{10}{*}{\footnotesize $n$}& \bf 1
&29.01&43.59&43.59&43.59&43.59&43.59&43.59&43.59&43.59&43.59&43.59\\    \cline{2-13}
&\bf 2&29.11&52.44&57.91&57.91&57.91&57.91&57.91&57.91&57.91&57.91&57.91\\    \cline{2-13}
&\bf 3&29.77&58.58&69.04&71.44&71.44&71.44&71.44&71.44&71.44&71.44&71.44\\    \cline{2-13}
&\bf 4&29.11&59.60&73.92&79.13&80.16&80.16&80.16&80.16&80.16&80.16&80.16\\    \cline{2-13}
&\bf 5&28.76&60.80&77.23&84.41&86.94&87.51&87.51&87.51&87.51&87.51&87.51\\    \cline{2-13}
&\bf 6&28.52&62.01&79.32&87.49&90.88&92.30&92.55&92.55&92.55&92.55&92.55\\    \cline{2-13}
&\bf 7&28.33&61.79&80.11&88.77&92.99&94.61&95.16&95.29&95.29&95.29&95.29\\    \cline{2-13}
&\bf 8&28.98&62.25&80.95&89.98&94.20&96.11&97.09&97.47&97.55&97.55&97.55\\    \cline{2-13}
&\bf 9&29.09&62.59&80.84&89.94&94.57&96.96&97.94&98.40&98.56&98.59&98.59\\    \cline{2-13}
&\bf 10&29.01&62.86&81.34&90.75&95.36&97.63&98.56&99.06&99.28&99.34&99.35\\    \cline{2-13}   
\hline
  \end{tabular}
  \caption{Approximated percentage value for $\ell=8$ and $m=8$.}
  \label{table88x}
\end{table}

\section{Conclusion}
We presented a way to get an approximated value for the number of non-equivalent \LFT{}s that are injective with some delay $\tau$. We also give a recurrence relation to determine the number of canonical \LFT{}s, and show how to get an approximated value for the percentage of equivalence classes formed by injective \LFT{}s. 

From the experimental
results presented in the previous section we may draw two very important conclusions. First, that the number of injective equivalence classes is very high and seems to grow exponentially as the structural parameters $\ell$ and $n$ increase. This implies that a brute force attack to the linear part of the key space may not be feasible. Second, that the percentage of
equivalence classes of $\omega$-injective \LFT{}s, with structural parameters $\ell,m,n$, is very high, for values of $n$ slightly larger than $\ell$ and $m$. The \LFT{}s used in Cryptography satisfy the condition $n=h\ell+km$, where $h,k \in \NN \setminus\{0\}$, which guarantees that $n$ is large enough so that there is a very high percentage of $\omega$-injective \LFT{}s of that size. Therefore, random generation of LFTs is a feasible option to generate keys. 

These results constitute an important step towards the evaluation of the key space. A similar study is required for the non-\LFT{}s used in the FAPKCs.
\section*{Acknowledgements}
This work was partially funded by the European Regional Development Fund through the programme COMPETE and by the Portuguese Government through the FCT under projects PEst-C/MAT/UI0144/2013 and  FCOMP-01-0124-FEDER-020486. Ivone Amorim is funded by the FCT grant SFRH/BD/84901/2012.

 The authors gratefully acknowledge the useful suggestions and comments of the anonymous referees.
 
\biblio{StatisticalStudyInvertibility}

\end{document}